\documentclass[a4paper,11pt]{article}
\usepackage{pos}
\usepackage{subfigure}

\title{Usage of GPUs for online and offline reconstruction in ALICE in Run 3}

\author*[a]{David Rohr for the ALICE Collaboration}

\affiliation[a]{CERN,\\
  1211 Geneva 2,\\
  Switzerland}

\emailAdd{drohr@cern.ch}

\abstract{In Run 3, ALICE records Pb--Pb collisions at an unprecedented rate of 50 kHz, storing all data in a continuous readout (triggerless) mode.
The main purpose of the ALICE online computing farm is the calibration of the detectors and the compression of the recorded data.
The detector with the largest data volume by far is the TPC, and the online farm is thus optimized for the fast and efficient processing of TPC data during data taking.
For this, ALICE leverages heavily the computing power of GPUs.
When there is no beam in the LHC, the GPU-equipped farm performs the offline reconstruction of the recorded data, in addition to the GRID.
Since the majority of the computing capacity of the farm is in the GPUs, and meanwhile also some GRID sites begin to offer GPU resources, ALICE has started to offload other parts of the offline reconstruction to GPUs as well.
The talk will present the experience and processing performance with GPUs in the Run 3 Pb--Pb and pp online and offline processing in ALICE.
}

\FullConference{42nd International Conference on High Energy Physics (ICHEP2024)\\
18-24 July 2024\\
Prague, Czech Republic\\}

\begin{document}
\maketitle

\section{Introduction}

ALICE (A Large Ion Collider Experiment) \cite{bib:alice} at the CERN LHC (Large Hadron Collider) was upgraded in the Long Shutdown 2 to operate in a continuous readout mode instead of triggered readout.
This enables ALICE to record heav- ion collisions at up to 50 kHz interaction rate in Run 3, which is two orders of magnitude more than in Run 2.
Since ALICE does not employ a trigger but stores all recorded heavy-ion data, the reduction of the data size to storage is one of the most important tasks during data taking.
Consequently, an update of the whole online computing infrastructure and the computing scheme was done in parallel to the detector upgrade \cite{bib:o2tdr}.
ALICE employs an online computing farm consisting of 350 servers each equipped with 8 GPUs called the EPN farm (Event Processing Nodes).
During data taking, the EPN farm performs detector calibration, full event reconstruction of a small subset of the collisions for visualization and quality control purposes, and most importantly data compression.
The detector with the largest data volume of ALICE is the TPC (Time Projection Chamber), thus compression of TPC data is critical, and ALICE employs several processing steps to reduce the TPC data size \cite{bib:ctd2019}.

ALICE employs the O$^2$ software framework, which was developed for the processing of Run~3, and which is a unified framework used for online processing, offline processing, simulations, and analysis \cite{bib:chep2023}.
The same software runs online and offline, albeit with a different setting to ensure the required real-time processing speed.

\begin{table}[b]
\centering
\caption{Relative processing times of the synchronous reconstruction steps processing 50 kHz Pb--Pb Monte-Carlo simulated data. Calibration and QC tasks are not shown in this table..}
\label{tab:syncreco}
\begin{tabular}{lr}
\hline
Processing step & Relative time \\
\hline
TPC Processing (Tracking, Clustering, Compression) & 99.37\% \\
EMCAL Processing & 0.20\% \\
ITS Processing (Clustering + Tracking) & 0.10\% \\
TPC rANS Encoding & 0.10\% \\
ITS--TPC matching & 0.09\% \\
MFT Processing & 0.02\% \\
TOF Processing and Global Matching & 0.02\% \\
\hline
Rest & 0.1\% \\
\hline
\end{tabular}
\end{table}

\looseness=-1
The driving factor for relying on GPUs for the online processing was the experience with the TPC processing on GPUs in Run 2 \cite{bib:hltpaper}, and the fact that the TPC, as the detector with the largest data volume, causes the majority of the online processing workload of Run 3.
In order to be vendor-independent and to support the largest variety of GPU models, all ALICE GPU code is written in a generic C++ language and can dispatch to hardware that supports CUDA, ROCm, or OpenCL \cite{bib:generic}.
The GPUs used in the EPN computing farm are AMD MI50 and MI100.
When there is no beam in the LHC, the EPN computing farm is used for offline processing.
Using the GPUs also for offline processing enables ALICE to leverage the computing power of the EPN farm, and will also achieve a better utilization of GPU-enabled GRID sites, thus ALICE is currently testing other GPU models, e.\,g. from NVIDIA, in the GRID.
To improve the GPU utilization in offline, ALICE aims to offload more and more reconstruction tasks to the GPUs in the future.

\section{Differences between online and offline processing}

While ALICE runs the same software online and offline, there is a huge difference in the selection of algorithms that are executed, in the calibration, and particularly in the fraction of the statistics processed by the different algorithms.
In the online processing, ALICE has to do full tracking for the TPC, but it runs tracking only for a small fraction of the data for the other detectors.
Thus, the processing is fully dominated by the TPC tracking which can run on GPUs.
In the offline processing, the workload distribution is much more heterogeneous.

\begin{table}[b]
\parbox{.48\linewidth}{
\centering
\caption{Relative processing times of the asynchronous reconstruction steps processing 650 kHz pp real data of 2022, no calorimeters in the run.}
\label{tab:asyncreco}
\begin{tabular}{lr}
\hline
Processing step & Relative time \\
\hline
TPC Processing (Tracking) & 61.41\% \\
ITS--TPC matching & 6.36\% \\
MCH & 6.13\% \\
TPC Entropy Decoding & 4.65\% \\
ITS Tracking & 4.16\% \\
TOF Matching & 4.12\% \\
TRD Tracking & 3.95\% \\
MCH Tracking & 2.02\% \\
AOD Production & 0.88\% \\
\hline
Quality Control & 4.00\% \\
\hline
Rest & 2.32\% \\
\hline
\end{tabular}
}
\hfill
\parbox{.48\linewidth}{
\centering
\caption{Relative processing times of the asynchronous reconstruction steps processing 47 kHz Pb--Pb real data of 2023.}
\label{tab:asyncrecopbpb}
\begin{tabular}{lr}
\hline
Processing step & Relative time \\
\hline
TPC Processing (Tracking) & 52.39\% \\
ITS Tracking & 12.65\% \\
Secondary Vertexing & 8.97\% \\
MCH & 5.28\% \\
TRD Tracking & 4.39\% \\
TOF Matching & 2.85\% \\
ITS--TPC Matching & 2.64\% \\
Entropy Decoding & 2.63\% \\
AOD Production & 1.72\% \\
\hline
Quality Control & 1.64\% \\
\hline
Rest & 4.84\% \\
\hline
\end{tabular}
}
\end{table}

\looseness=-1
Tables \ref{tab:syncreco}, \ref{tab:asyncreco}, and \ref{tab:asyncrecopbpb} show the relative computing time of the running tasks in an online environment compared to the offline processing of pp and Pb--Pb data.
Only one online processing case is shown, since they are all very similar and the TPC processing will always be above 98\%.
Also note that the online table shows only reconstruction tasks, and leaves out quality control, calibration, event building, and network IO.
Compared to the total CPU workload including these tasks, the TPC processing is still always above 90\% in the online case.
The EPN farm was designed such that 90\% of the computing power is provided by GPUs, thus the online processing can and will always only be GPU-bound.
Consequently, the EPN farm was dimensioned with sufficiently many GPUs to handle the highest computing workload at 50 kHz Pb--Pb interaction rate including a safety margin of 30\%.

As Tables \ref{tab:asyncreco} and \ref{tab:asyncrecopbpb} show, the workload during the asynchronous processing is more heterogeneous.
The TPC processing is still dominant, but it constitutes only 50\% to 60\% of the total workload, so the other tasks are not negligible.
ALICE has been using the GPUs for offline processing on the EPNs since 2023, offloading the TPC reconstruction.
The achieved speedup reaches quite exactly the expectation that offloading 50\% or 60\% should achieve a speedup of $2\times$ or $2.5\times$ on the EPN~\cite{bib:chep2023}.
Here, the processing on the EPN is CPU-bound, in contrast to the online processing which is GPU-bound, as the GPUs have 90\% of the total computing power when running highly optimized TPC reconstruction.
Other algorithms might experience a different speedup when ported from CPU to GPU.
Overall, It should thus be possible to offload up to 90\% of the offline tasks to the GPUs, before the limitation will switch from the CPU to GPU.
In particular, since offline processing is currently CPU-bound, it makes sense to offload the tasks to the GPU even in an inefficient fashion, as long as this frees up some CPU resources.
Thus, the expected fraction that can be offloaded to the GPU before running into a GPU-limit will be less than 90\% for the offline processing.
Currently, ALICE aims to offload the full central-barrel global tracking chain to GPUs, which includes the ITS, TPC, TRD, and TOF tracking and matching as well as secondary vertexing and track refit.
In both presented offline cases in Tables \ref{tab:asyncreco} and \ref{tab:asyncrecopbpb}, this amounts to 80\% of the total workload, which should yield close to optimal GPU utilization.
This is a gradual improvement, and a recent addition was the porting of TPC-track-model decoding to GPU in 2024, which sped up the offline reconstruction with GPUs by 1.2\% to 3\% \cite{bib:gabrielelhcp}.

\section{Scheduling}

Comparing Tables \ref{tab:asyncreco} and \ref{tab:asyncrecopbpb}, it can be seen that not only the TPC fraction, which is offloaded to GPU, is different, but also the relative fraction of the tasks currently running on CPU.
This adds additional complexity when aiming for a workflow that can process all data sets efficiently.
In case individual tasks are slower than the GPU processing, the ALICE Data Processing Layer (DPL) runs multiple instances of the tasks, processing data in a round-robin fashion to increase the throughput \cite{bib:chep2023}.
For this, the workflow specification needs to configure the process multiplicities explicitly.
But when the relative performances change, also the required multiplicities change, which happens for different particle species, different interaction rates, but also different software versions or configurations and calibrations.
For instance the tracking in the ITS is quite sensitive to the local multiplicity, thus it takes about 3 times more time in Pb--Pb compared to pp.
Unfortunately, it is not possible to simply run all tasks with high multiplicity, since the tasks are independent operating system processes and the available memory limits how many tasks can run in parallel.
Therefore, a heuristic algorithm guesses the best multiplicities, and this algorithm must be tuned regularly.

Another difference between the online and offline is the publishing of data at the source of the processing chain.
Technically, this is quite simple in the online: data is published at the source of the processing chain when it is recorded from the detector at a more or less fixed rate, with data rates depending on the interaction rate.
The processing must simply keep up with the data taking, while preserving a certain safety margin.
For ALICE this means having enough GPUs to process the highest expected input rate of 50 kHz Pb--Pb plus the safety margin of 30\% on GPUs, and having a slightly larger safety margin for the CPUs to ensure the processing is always GPU-bound.
In the offline processing, it would be desirable to achieve maximum CPU and GPU utilization.
But since the current ALICE offline processing is inevitably CPU-bound, ALICE aims to have at least maximum CPU utilization.
This means data must be published fast enough to keep the CPUs fully busy, but not too fast to avoid that too much data is in flight which could exhaust the memory.
DPL provides multiple mechanism to avoid that too much data is in flight.

\begin{figure}[t!]
\centering
\begin{subfigure}
\centering
\includegraphics[width=300px]{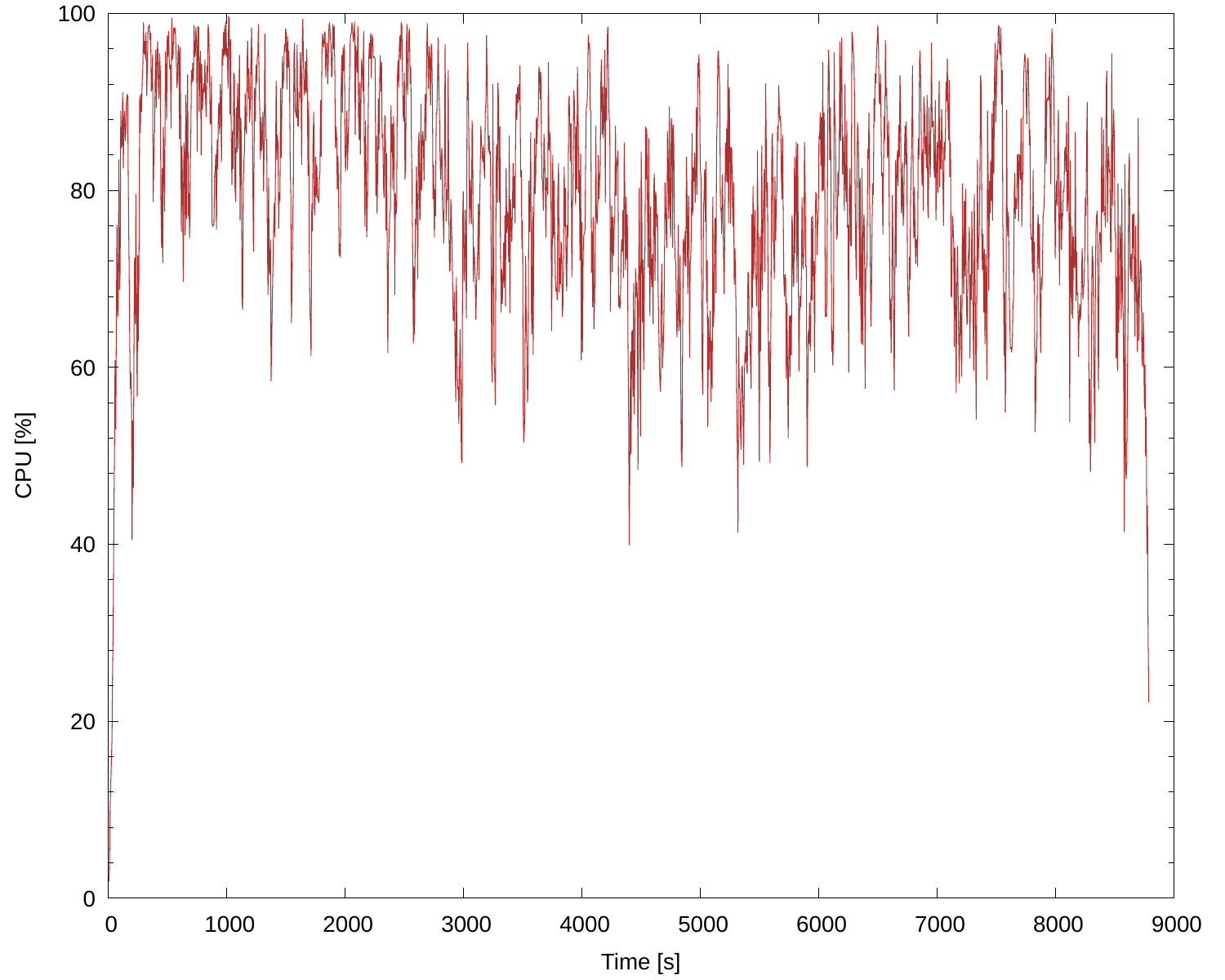}
\caption{CPU utilization during asynchronous processing on the EPN publishing as fast as possible with rate limiting.}
\label{fig:nosmooth}
\end{subfigure}
\begin{subfigure}
\centering
\includegraphics[width=300px]{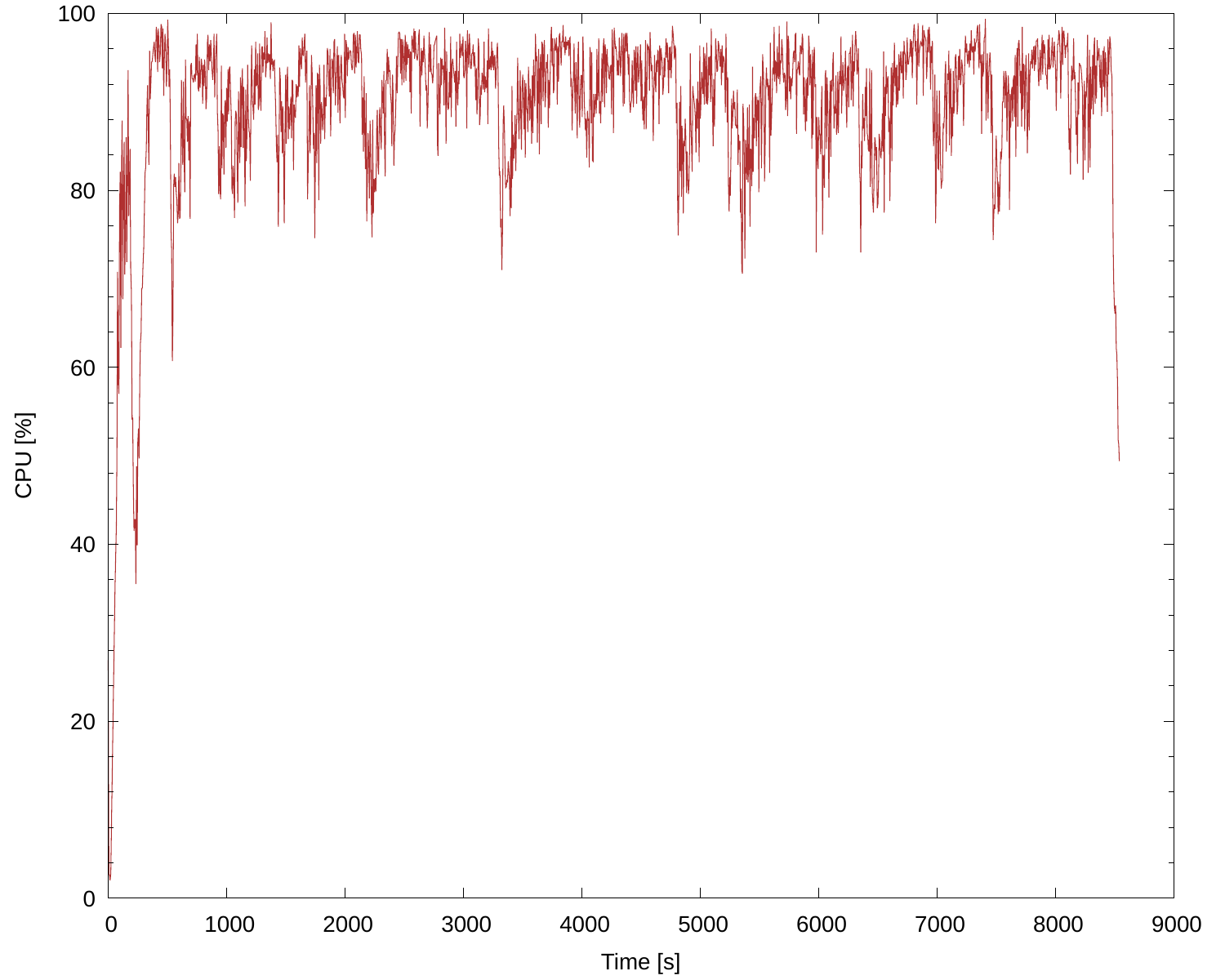}
\caption{CPU utilization during asynchronous processing on the EPN with publishing rate smoothing.}
\label{fig:smooth}
\end{subfigure}
\end{figure}

A naive rate-limiting approach publishes data as fast as possible at the source, and simply counts how many collisions are in flight, throttling the source when a limit is reached, until the first have reached the end of the chain.
Unfortunately, this yields oscillations when waves percolate through the processing graph.
When such a wave hits a bottleneck in the graph, for instance a fast but serial process with low process multiplicity, not all data can be processed in parallel, leading to a drop in CPU utilization.
Figure~\ref{fig:nosmooth} shows the resulting CPU usage over an offline computing job, which shows significant drops in the utilization.
Therefore, ALICE uses a heuristics to smoothen the publishing rate, aiming for homogeneous horizontal and vertical parallelization \cite{bib:chep2023} in the DPL processing graph.
Figure~\ref{fig:smooth} shows the resulting CPU usage, which is improved significantly.
The heuristics constantly tries to increase the CPU usage by gradually, slowly decreasing the publishing delay to have more parallel collisions in flight, which at some point causes back pressure and a decrease of the processing rate.
This regular pattern is clearly visible, and could probably still be improved.
Overall, the average CPU utilization is above 90\%.
Considering that the cores provide HyperThreading, this means that all physical cores are used and 80\% of the hyperthreaded ones.
Because the HyperThreading only adds ~30\% of additional computing power to the physical cores, more than 95\% of the available CPU computing capacity is used at the moment.

\section{Conclusions}

After the good experience of GPU usage in Runs 1 and 2, ALICE is relying heavily on GPUs to speed up the online and offline processing in Run 3.
Currently, 99\% of the online reconstruction workload runs on GPU, with no reason to offload the rest as this would complicate things for no benefit.
Since 2023, ALICE uses the GPUs on the EPN farm for offline processing offloading up to 60\% achieving a speedup of $2.5\times$ and of $2\times$ for pp and for Pb--Pb data.
The aim is to increase the offloaded fraction for both pp and Pb--Pb to 80\% for an expected total speedup of $5\times$.
ALICE will use GPUs at GRID sites in the future as well, and the GPU framework supports all GPUs that can run with CUDA, ROCm, or OpenCL.

\looseness=-1
The usage of the GPUs for online processing for both pp and Pb--Pb was smooth in 2022 and 2023, and the EPN farm could handle the highest rate observed so far of 47 kHz Pb--Pb collisions in 2023.
A hypothetical CPU-only online farm would require over 3000 servers with 64 physical cores each which would be prohibitively expensive.
Therefore, the GPU-enabled online-processing farm is an essential feature, which enables ALICE to run data taking at the highest rates in Run 3.

\end{document}